Micron 176 (2024) 103557

Contents lists available at ScienceDirect

# Micron

journal homepage: www.elsevier.com/locate/micron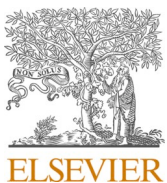
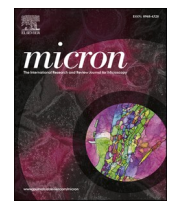

# Quantifying colors at micrometer scale by colorimetric microscopy (C-Microscopy) approach

Benedykt R. Jany

*Marian Smoluchowski Institute of Physics, Faculty of Physics, Astronomy and Applied Computer Science, Jagiellonian University, Lojasiewicza 11, 30348 Krakow, Poland*ARTICLE INFO

*Keywords:*
Microscopy
Hyperspectral Reflectance
Color Science
Colorimetric Microscopy
C-MicroscopyABSTRACT

The color is the primal property of the objects around us and is direct manifestation of light-matter interactions. The color information is used in many different fields of science, technology and industry to investigate material properties or for identification of concentrations of substances. Usually the color information is used as a global parameter in a macro scale. To quantitatively measure color information in micro scale one needs to use dedicated microscope spectrophotometers or specialized micro-reflectance setups. Here, the Colorimetric Microscopy (C-Microscopy) approach based on digital optical microscopy and a free software is presented. The C-Microscopy approach uses color calibrated image and colorimetric calculations to obtain physically meaningful quantities i. e., dominant wavelength and excitation purity maps at micro level scale. This allows for the discovery of the local color details of samples surfaces. Later, to fully characterize the optical properties, the hyperspectral reflectance data at micro scale (reflectance as a function of wavelength for a each point) are colorimetrically recovered. The C-Microscopy approach was successfully applied to various types of samples i.e., two metamorphic rocks unakite and lapis lazuli, which are mixtures of different minerals; and to the surface of gold 99.999 % pellet, which exhibits different types of surface features. The C-Microscopy approach could be used to quantify the local optical properties changes of various materials at microscale in an accessible way. The approach is freely available as a set of python jupyter notebooks.## 1. Introduction

The color is the primal property of the objects around us. The color which we see is a direct manifestation of light-matter interactions (Rivera et al., 2020). Incident photons interact with matter via several different complicated processes which in consequence, simplifying it, leads to the absorption of some of them, rest is reflected and comes to the detector system (i.e. our eyes or digital camera) which is later recorded as particular color value. The color as defined by the by the International Commission on Illumination (CIE) in 1931 as tristimulus color values $J$ : $[X, Y, Z]$ are expressed as (C.I.E Report, 2004; Janos Schanda, 2007): $J = \frac{1}{N}\int_\lambda R(\lambda)I(\lambda)\bar{j}(\lambda)d\lambda$ and $N = \int_\lambda \bar{y}(\lambda)d\lambda$ for $\bar{j}(\lambda) : [\bar{x}(\lambda), \bar{y}(\lambda), \bar{z}(\lambda)]$, where $R(\lambda)$ is a spectral reflectance, $I(\lambda)$ is spectral power distribution of light source (illuminant). The integral is computed over the visible light wavelength $\lambda$ range. Here $\bar{j}(\lambda)$ denotes CIE standard observer matching functions, which describe the spectral sensitivity (chromatic response) of the standard observer. In case of the commonly used RGB sensor ($J : [R, G, B]$) this looks very similar with $\bar{j}(\lambda)$ being spectral sensitivity of the j-th channel (Lin et al., 2023). From the above one can see that there is a direct correspondence between the color and the optical properties expressed as spectral reflectance $R(\lambda)$, which is related to the refractive index of the material (Hummel, 2011) $n(\lambda)$ via $R(\lambda) = \frac{(n(\lambda)-1)^2 + k(\lambda)^2}{(n(\lambda)+1)^2 + k(\lambda)^2}$, where $k(\lambda)$ is the extinction coefficient, and is related to the complex dielectric function (Hummel, 2011) $\epsilon(\lambda) = [n(\lambda) - ik(\lambda)]^2$. Since the color is the consequence of the basic interactions it could be calculated from first principles by quantum mechanical calculations using Density Functional Theory (DFT) (Prandini et al., 2019). The DFT approach can be used to successfully predict the color of various metals (Prandini et al., 2019). Experimentally the information about the color is measured by colorimeters or spectrophotometers and it is used in different fields of science and technology e.g., for measurements of nutrients (Kim et al., 2022), detection of metal ions (Kalluri et al., 2009), in materials science for coatings design (Chen et al., 2020) and also in heritage science for work conservation (Striova et al., 2018). Colorimetric information registered by smartphone was also used to detect anemia in infants and young children without taking blood samples

*E-mail address:* benedykt.jany@uj.edu.pl.

https://doi.org/10.1016/j.micron.2023.103557
Received 15 May 2023; Received in revised form 18 September 2023; Accepted 12 October 2023
Available online 14 October 20230968-4328/© 2023 The Author(s). Published by Elsevier Ltd. This is an open access article under the CC BY license (http://creativecommons.org/licenses/by/4.0/).

(Wemyss et al., 2023). In many of these research fields the color information is employed as a global property in macro scale.

To characterize qualitatively local color changes in micro scale, which is important in materials research (Papadopoulos et al., 2018), photonic crystals (Sowade et al., 2016) and 2D material (Frisenda et al., 2017) characterization, one has to use dedicated microscope spectrophotometers or specialized micro-reflectance setups.

Recently also several approaches utilizing color information at nano and micro scale were developed like: colorimetric histology (Balaur et al., 2021a) which utilizes special plasmonically active microscope slide to enhance color of biological samples without the need of dyeing or staining; dedicated Gires-Tournois Immunoassay Platform (GTIP) (Yoo et al., 2022) for label-free bright-field imaging of nanoscale bioparticles; special combination of plasmonically active metamaterials together with ptychographic coherent diffractive imaging (Balaur et al., 2021b) which yields additional contrast and enables imaging extremely thin and transparent objects like thin tissue sections.

Here, the Colorimetric Microscopy (C-Microscopy) approach based on commonly available digital optical microscopy and a free software is introduced. The C-Microscopy approach uses color calibrated (D65 illuminant) image and colorimetric calculations to obtain physically meaningful quantities i.e. dominant wavelength and excitation purity maps at micro level scale. The color details are uncovered, the microscopic images are colorimetrically characterized in a quantitative way at a micrometer level. To fully characterize the optical properties, the spectral reflectance (reflectance as a function of wavelength) for a each point is colorimetrically recovered forming hyperspectral reflectance data. This allows for the full quantification of the local optical properties changes at micrometer scale.

## 2. The idea of the colorimetric microscopy (C-Microscopy) approach

The idea of the the approach of the Colorimetric Microscopy (C-Microscopy) based on digital optical microscopy and a free software is graphically presented in Fig. 1. The approach consists of the following steps:

1. In the first step, a microscopic image is collected using digital optical microscope in reflected light mode as a color RGB image. Here, a digital optical microscope Delta Optical Smart 5MP PRO was used. The microscope has a 5MP CMOS sensor together with an illumination system consisting of 8 ultra bright white LEDs with smooth illumination intensity adjustment and a dedicated optical system of lenses, which moves relative to the sensor. The microscope is coupled to the tripod which allows for the height adjustments. The data were collected using free cross-platform software AstroDMx Capture 1.7.1.0 (https://www.astrodmx-capture.org.uk/) under the Linux Mint 20.3 operating system running on PC. The data were collected as a series of sixteen 8-bit RGB tiff images each with a resolution of $2592 \times 1944$ pixels, which were later median stacked. The data were captured in the manual white balance mode with 15 ms exposure time for each frame. After data collection image of a length calibration slide, provided by microscope manufacturer, was also recorded.

2. In the next step, the microscope was precisely color calibrated (D65 illuminant). Here color calibration chart Calibrite ColorChecker Classic Mini was used. The calibration chart consists of 24 different color references, as presented in Fig. 2a). Each of the 24 colors was measured under the microscope using exactly the same conditions as for the sample measurements. The observed color values are presented in Fig. 2b). The comparison between reference color values and observed for each of 24 colors for red, green and blue channel is presented in Fig. 2d)-f). Its is seen that the observed values do not lie on the straight line (perfect agreement). The calibration was performed using IJP-Color plugin (https://github.com/ij-plugins/ijp-color) for free software ImageJ/FIJI (Schindelin et al., 2012). The calibration was performed using reference color values for D65 illuminant. The calibration was performed for each channel (red (r), green (g), blue (b)) using third order polynomial of the form constant $+ r + g + b + r*r + r*g + r*b + g*g + g*b + b*b + r*r*r + r*r*g*b + g*g*g + b*b*b$, "cubic cross-band" as in the IJP-Plugin, which takes into account correlations between the channels up to the third order. The obtained calibration coefficients are presented in Table S1 in Supporting Information. The performance of the calibration was evaluated by plotting corrected color values after calibration versus the reference color Fig. 2d)-f). It is seen that now all the colors lie on the straight line (perfect match). To evaluate the uncertainty of the performed calibration absolute color value difference between corrected and reference color versus reference color was plotted for each channel (red, green blue) Fig. 2g). Average difference of $\Delta R = 3.95$, $\Delta G = 2.16$ and $\Delta B = 3.71$ for red, green and blue channel respectively was obtained. This defines the average systematic uncertainty for each channel related to the performed color calibration. The average ΔRGB= 6.78 was obtained, corresponding to the relative average ΔRGB= 1.78%, which expresses relative average precision of the performed color calibration. Exemplary data and ImageJ/FIJI script for color chart creation from measurements are provided for the color calibration step. The data are available from Zenodo (Benedykt R. Jany, 2023) (https://doi.org/10.5281/zenodo.7789585). The performed calibration was applied to the collected microscopic images. Finally each image is color calibrated (D65 illuminant). This guarantees that the colors of the samples at microscale after calibration are reliably reproduced always in the same way independently on the initial microscope conditions.

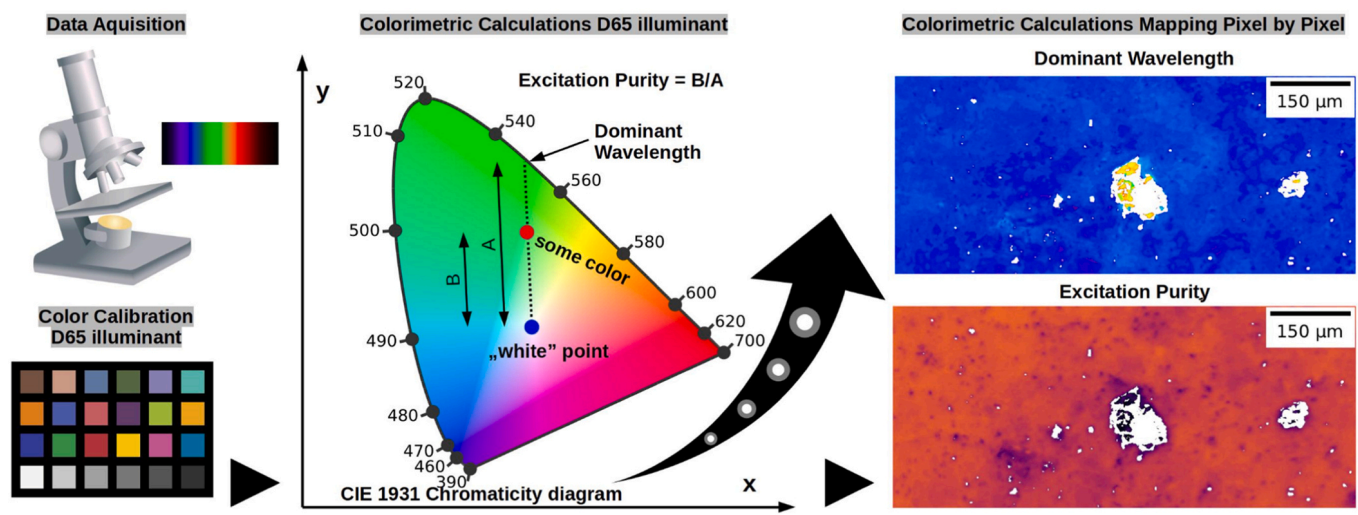

**Fig. 1.** Graphical representation of the idea of the colorimetric microscopy (C-Microscopy) approach. The data are acquired using digital optical microscope in the form of color RGB image. The microscope is color calibrated (D65 illuminant) using color calibration chart. Next, the colorimetric calculations are performed based on CIE 1931 chromaticity diagram (D65 illuminant) to determine the dominant wavelength and excitation purity of the color. The calculations are performed pixel by pixel to form a map of a dominant wavelength and excitation purity. The color details are uncovered, the microscopic image of the sample is colorimetrically characterized in a quantitative way at a micrometer level. Later, also hyperspectral reflectance is colorimetrically recovered, see details later in the text.

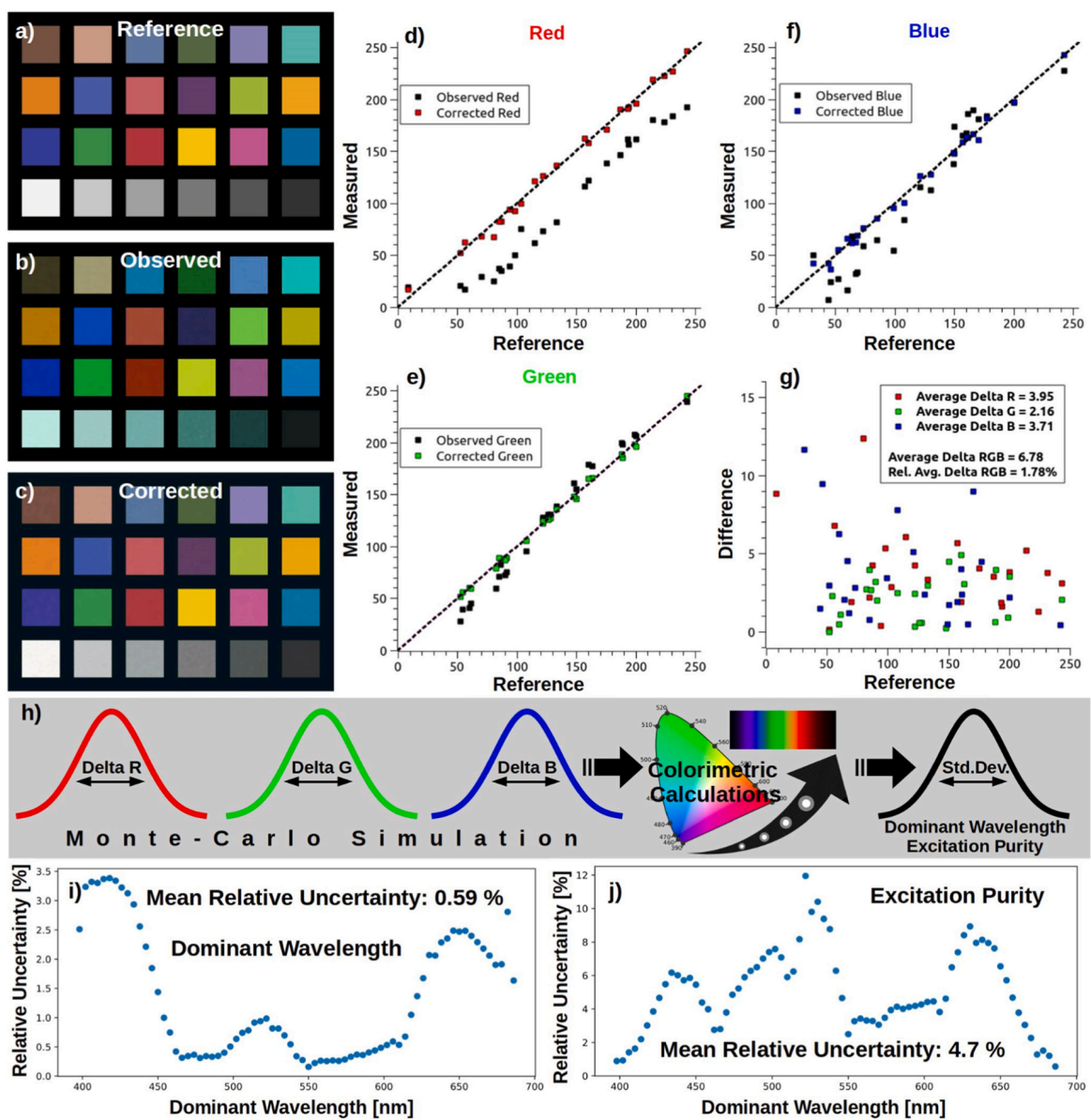

**Fig. 2.** Color calibration (D65 illuminant) of microscope. Reference a) observed b) and corrected c) colors of used calibration chart (Calibrite ColorChecker Classic Mini). Measured versus reference color value for the Red d), Green e) and Blue f) channel. Observed and corrected color values indicated. Absolute color value difference between corrected and reference color versus reference color for each channel (red, green blue) g). Relative average precision of color calibration of 1.78 % is achieved, expressed as Delta RGB. Schematic of systematic uncertainty calculation using Monte-Carlo simulation for dominant wavelength and excitation purity related to color calibration h). Obtained relative systematic uncertainties, related to color calibration, for dominant wavelength i) and excitation purity j) as a function of dominant wavelength. Mean relative systematic uncertainty of 0.59 % is obtained for dominant wavelength and 4.7% for excitation purity.

3. In the following step, the colorimetric calculations are performed to extract from the D65 calibrated images physically meaningful quantities. Each RGB pixel on the image is first converted to the CIE 1931 xy chromaticity coordinates. Next dominant wavelength and excitation purity is calculated for each pixel (C.I.E. Report, 2004; Hunt, 2011), see Fig. 1. The dominant wavelength is defined as a cross section point at the spectral locus border of the chromaticity diagram of the line between "some color" and a "white point" on the diagram. The dominant wavelength of a "some color" is a wavelengths of a monochromatic light which in consequence gives color appearance of the same hue as a "some color". This might be also interpreted as a dominant reflected wavelength from the sample surface. The excitation purity is defined as ratio of the distance between "white point" and "some color" to the distance between "white point" and dominant wavelength point on the chromaticity diagram. The excitation purity describes how pure is the color i.e. for the pure colors on the spectral locus the excitation purity is equal to 1, when the color approaches towards the "white point" (is more and more mixed with white) the excitation purity decreases. The dominant wavelength and excitation purity are colorimetric quantities (not spectroscopic) which describe the color in quantitative way. The calculations are performed pixel by pixel to form a maps of a dominant wavelength and excitation purity and their histograms. The correlation plot between excitation purity and dominanat wavelngth is also calculated. The dominant wavelength construction cannot be performed for the colors lying between the violet end and red end on the diagram. These colors are mixtures of red and blue, there does not exist single wavelength of monochromatic light to describe them. These areas are indicated on the maps as white color. This finally allows for the uncovering of color details, not visible directly on the microscopic images, which results in the colorimetric characterization and interpretation in a physical meaningful quantitative way the collected microscopic images of the samples. Later, also hyperspectral reflectance, where the reflectance R in measured for each x, y pixel position on the image, is colorimetrically recovered from measured calibrated color values, see details later in the text.

4. In the final step, the systematic uncertainties related to the performed color calibration are propagated to the calculated colorimetric quantities i.e. dominant wavelength and excitation purity. The uncertainty propagation is performed using Monte-Carlo method, see Fig. 2h. Three Gaussian distributions with sigma $\Delta R$ = 3.95, $\Delta G$ = 2.16 and $\Delta B$ = 3.71 (for red, green and blue channel respectively) and mean taken from uniform distribution (100,000 points), for each Gaussian distribution 1000 points was generated. For each generated point in RGB space colorimetric calculations were performed, dominant wavelength and excitation purity were calculated. Next, for each generated mean RGB value represented as three dimensional Gaussian distribution in RGB space, relative dominant wavelength systematic uncertainty and relative excitation purity systematic uncertainty were calculated as a ratio of standard deviation to mean value of the resulted dominant wavelength and excitation purity distribution. The calculated systematic uncertainties are presented as a function of dominant wavelength in Fig. 2i)-j). It is seen that relative systematic uncertainty of the dominant wavelength Fig. 2i) changes with dominant wavelength from ~0.3 % to ~3.5 % and relative systematic uncertainty of the excitation purity Fig. 2j) changes with dominant wavelength from ~1 % to ~12 %. The mean relative systematic uncertainty of 0.59 % is obtained for dominant wavelength and 4.7 % for excitation purity. This shows that dominant wavelength and excitation purity values could be extracted by this approach in a reliable way, as indicated by the uncertainties.

All the colorimetric calculations are performed using D65 illuminant and CIE 2 deg standard observer which mimics the illumination of the microscope (Gutiérrez, 2021). The colorimetric calculations are implemented in python using freely available software libraries as a building blocks like Numpy (Walt et al., 2011) and Color Science (Mansencal et al., 2022) . All calculations were performed on the standard desktop PC running Linux Mint 20.3 operating system. The program uses as an input D65 calibrated RGB image (tiff, png, jpg) and outputs for the convenience report file in the PDF format with calculated maps and distributions. The maps are also saved as a tiff files for later processing e. g. in ImageJ/FIJI (Schindelin et al., 2012). The python jupyter notebooks, together with exemplary data, which perform colorimetric calculations, Monte-Carlo uncertainties propagation and colorimetric recovery of hyperspectral reflectance are freely available from Zenodo (Benedykt R. Jany, 2023) (https://doi.org/10.5281/zenodo.7789585).

## 3. Applications of C-Microscopy approach to different samples

The approach of C-Microscopy was successfully applied to various types of samples with different colors like two metamorphic rocks green-pink unakite and blue-yellow lapis lazuli, which are mixtures of different minerals. The approach was also applied to the analysis of the surface of gold 99.999% pellet, which exhibits different types of surface features.

### 3.1. Metamorphic rocks unakite and lapis lazuli

The C-Microscopy approach was applied to two samples of metamorphic rocks unakite and lapis lazuli in the form of polished plates (see Fig. 3). Unakite is a metamorphic rock, a type of granite consisting mainly of three different minerals green epidote, pink orthoclase feldspar and white quartz (Franz and Liebscher, 2004). When polished it is used as a semiprecious stone. Lapis lazuli is also a metamorphic rock it consists mainly of the following minerals: blue lazurite, yellow pyrite and white calcite (Colomban, 2014). Lapis lazuli when polished is used in jewelry. After grinding lapis lazuli to a powder it is used as a natural pigment ultramarine. Ultramarine pigment during the Renaissance was the most expensive blue pigment used by the painters (Emerson, 2015). The macro scale view of the measured samples is presented in Supporting Information in Fig. S7. Fig. 3a) shows optical microscopy image, color calibrated (D65 illuminant), of the Unakite surface. From the optical microscopy image itself it is not so obvious where are the regions of the three different minerals and how they merge into another. After applying C-Microscopy approach to the calibrated image the maps of dominant wavelength Fig. 3b) and excitation purity Fig. 3c) together with their distributions Fig. 3d)-e) are derived. The dominant wavelength clearly shows two regions: yellow green region related to the green epidote and orange-red region related to the ping orthoclase feldspar Fig. 3b). The white quartz is mixed into epidote and feldspar, changing its color purity, which is clearly visible on the excitation purity maps Fig. 3c) as the regions of low intensity. The dominant wavelength changes from around 560 nm to ~680 nm, indicating strong light absorption in other areas of visible spectrum. The mean dominant wavelength (averaged over whole microscopic image) of the Unakite is 586.6 nm and the mean excitation purity is 0.375, Fig. 2d)-e). The correlation between excitation purity and dominant wavelength is also calculated in the C-Microscopy approach, see Fig. S1 in the Supporting Information. From the correlation one can indicate the "purest dominant wavelength" (the dominant wavelength for the highest excitation purity) for the sample which is ~590 nm with an average excitation purity of ~0.4 (averaged over whole microscopic image).

Another example shows optical microscopy image, color calibrated (D65 illuminant), of the Lapis Lazuli surface Fig. 3f). Applying C-Microscopy approach to the calibrated image the maps of dominant wavelength Fig. 3g) and excitation purity Fig. 3h) together with their distributions Fig. 3i)-j) are derived. The dominant wavelength clearly shows the pyrite region as a yellow wavelength and lazurite as blue region wavelength. In the blue lazurite region in the dominant wavelength Fig. 3g) a structure is visible (light blue and dark blue regions) due to the appearance of white calcite. In the excitation purity map this is visible as a darker regions Fig. 3g). Combining information from dominant wavelength map and excitation purity map one can easily indicate regions of pure blue lazurite. The dominant wavelength here changes from ~380 nm to ~490 nm (excluding pyrite region), this indicates strong light absorption in other areas of visible spectrum. The mean dominant wavelength (averaged over whole microscopic image) of the Lapis Lazuli is 448 nm and the mean excitation purity is 0.527, Fig. 3i)-j). The correlation between excitation purity and dominant wavelength, see Fig. S2 in the Supporting Information, shows the "purest dominant wavelength" (the dominant wavelength for the highest excitation purity) for the sample at ~550 nm with an average excitation purity of ~0.55 (averaged over whole microscopic image).

The measured by C-Microscopy at microscale color of Lapis Lazuli in CIE 1931 xy chromaticity coordinates (D65 illuminant) was compared with color of Mineral Lazurite Powder and artificial blue pigments (Ultramarine, Prussian Blue, Cobalt Blue), see Table S2 and details in Supporting Information. The color value of Lapis Lazuli, expressed as xy chromaticity coordinates (D65 illuminant), is significantly different from color of artificial paint pigments. This allows for the differentiation between the natural Lapis Lazuli and artificial blue pigments, thus making it possible to use C-Microscopy approach in the field of heritage science.

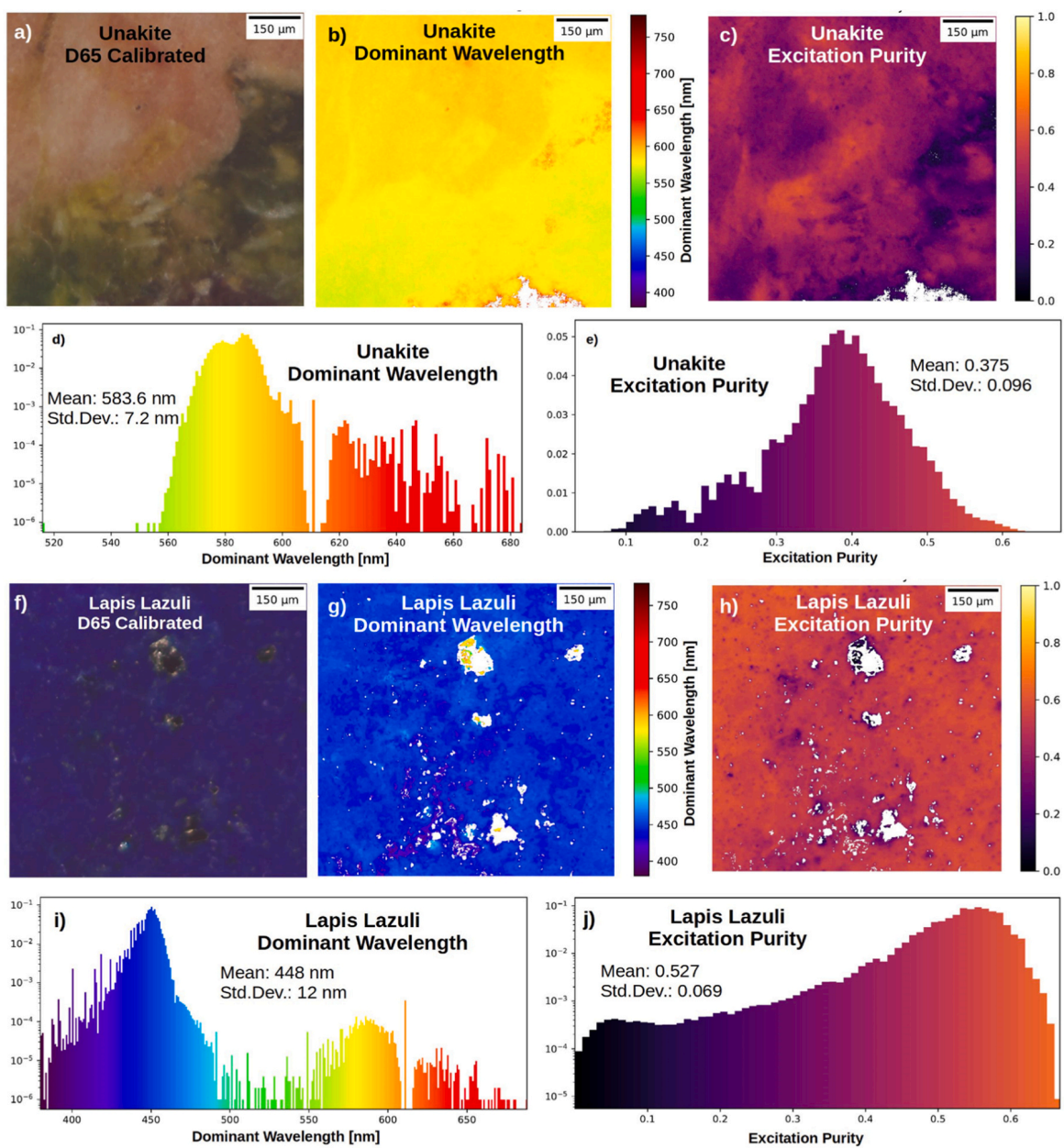

**Fig. 3.** Results of colorimetric microscopy (C-Microscopy) imaging of Unakite and Lapis Lazuli polished plates. Color calibrated (D65 illuminant) optical microscopy image of Unakite a) and Lapis Lazuli f). Dominant wavelength map of Unakite b) and Lapis Lazuli g). The regions of green epidote and pink orthoclase feldspar for the Unakite and the pyrite (yellow) inclusions for Lapis Lazuli are clearly visible. Excitation purity map of Unakite c) and Lapis Lazuli h). Local changes at micrometer level of dominant wavelength and excitation purity are clearly visible on the maps for both samples. Distributions of dominant wavelength and excitation purity for Unakite d), e) and Lapis Lazuli i), j). The Unakite dominant wavelength peaks at 583.6 nm while the Lapis Lazuli at 448 nm. The samples were colorimetrically characterized at micrometer level.

The samples of Unakite and Lapis Lazuli were colorimetrically characterized in a physical interpretable way at micrometer level.

### 3.2. Surface of gold 99.999 % pellet

The C-Microscopy approach was next applied to the surface of gold 99.999 % pellet (Kurt J. Lesker Company), see Fig. 4. Gold is a noble metal widely used in different fields of science and technology (Blumenstein, 2011; Jany, 2017; Sierant, 2021) but also in everyday use as a jewelry (Brody, 2013). It is also exceptionally interesting to study gold optical properties since recent studies on gold surface show that light interaction with metal surface is much more complex than thought (Strait, 2019). The light causes to move the electrons in the gold in the same or in the opposite direction as the photons (Strait, 2019). There is a speculation that the light may interact not only with free electrons in the metal but also with the core electrons (Buchanan, 2019). Gold is also chemically inert, it does not change over time, so it is an ideal "gold standard" of the color. Fig. 4a) shows color calibrated (D65 illuminant) optical microscopy image of gold surface. The average color of gold was calculated in CIE 1931 xy chromaticity coordinates (D65 illuminant), see Table 1. The measured average color of gold agrees with a theoretical gold color value (McPeak et al., 2015; Manas, 2020) as well as the color value from measured gold reflectance (Bass, 1995) within the range of calculated uncertainties. This validates the correctness of the presented C-Microscopy approach. As one can see from the optical microscopy image Fig. 4a), the surface of gold is not homogeneous in color. It exhibits different types of grooves on the surface. To visualize better this effects the optical microscopy image is presented as 3D surface plot, see Fig. 4b). The image luminance is interpreted as a height in the plot (Barthel, 2006; Surface-plot-3D).

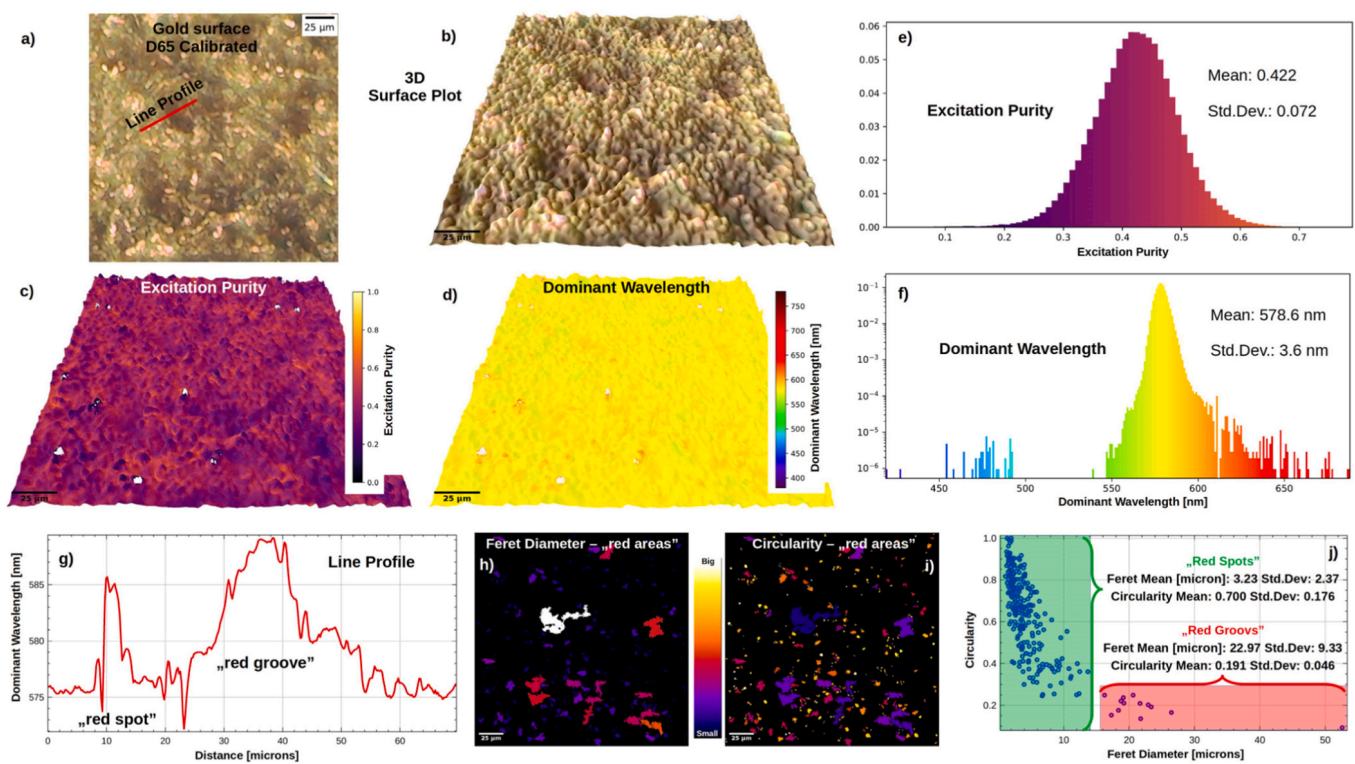

**Fig. 4.** Results of colorimetric microscopy (C-Microscopy) imaging of the surface of Gold 99.999% pellet. Color calibrated (D65 illuminant) optical microscopy image of Gold surface a). 3D surface plot image of Gold surface b). Different type of grooves are visible on the surface. Excitation purity and dominant wavelength maps c), d). Local changes of the color are visible on the surface. The area of the grooves exhibits dominant wavelength shifted to the red. This is effect is not directly visible on microscopic image itself. Excitation purity and dominant wavelength distributions e), f). Dominant wavelength line profile through the "red area" g), as indicated in a) by red line. Particle analysis resulted shape descriptors color coded maps of feret diameter h) and circularity i) (4 *π * Area/perimeter$^2$) for the "red area". Particles analysis resulted plot of circularity versus feret diameter for the "red area". It is seen that the shape of the "red area" (circularity) depends on the size (feret diameter). The surface of the gold is colorimetrically characterized at micrometer level.

**Table 1**
Average color of Gold in CIE 1931 xy chromaticity coordinates (D65 illuminant). The measured color of the gold in this work (Au 99.999% Pellet) agrees with a theoretical gold color value as well as the color value form measured gold reflectance within the range of calculated uncertainties.

| Sample Description | Average Gold Color CIE 1931 xy coordinates D65 illuminant |
| --- | --- |
| This Work (Au 99.999 % Pellet) | x: 0.392 ± 0.012 y: 0.400 ± 0.013 (95% CL) |
| Theoretical value (McPeak et al., 2015; Manas, 2020) | x: 0.38 y: 0.40 |
| From measured gold reflectance (Bass, 1995) | x: 0.3895 y: 0.4062 |

Looking into excitation purity and dominant wavelength C-Microscopy maps Fig. c)-d) one can see that the area of the grooves exhibits dominant wavelength shifted towards the red and small excitation purity. The changes of excitation purity with the dominant wavelength are clearly visible on the correlation plot Fig. S3 in Supporting Information. The effect of the red shift in the grooves is not directly seen on the microscopic image itself. Recently, the effect of red shift of the color in the grooves on the surface of gold, caused by the multiple reflections, was attributed to solve the puzzle of gold color variation towards the red color (Manas, 2020). The grooves look like black on the images but actually they are deep dark orange/red. The observed gold color inhomogeneity was attributed for a long time to variety of different factors like: "sample effects" (Aspnes et al., 1980; Winsemius et al., 1975) and even "gold rust" (Corregidor et al., 2013; Gusmano et al., 2004), but finally explained by the "structural color" caused by the grooves (Manas, 2020). Here, these "red grooves" are directly visible by the C-Microscopy approach at microscale. From the histograms Fig. 4e)-f) one can see that the average dominant wavelength is 578.6 nm and the average excitation purity is 0.422. Additionally to the "red grooves" on the sample surface much smaller and brighter "red spots" are also visible. The line profile through the typical "red area" in dominant wavelength is presented in Fig. 4g), as indicated in Fig. 4a) by red line. It is seen that the dominant wavelength changes smoothly from low value to high (in the center) for the "red groove", while for the "red spot" the changes are very rapid. Next, particle analysis (Schindelin et al., 2012) was performed on the "red areas", extracted by segmentation using thresholding dominant wavelength map above 583 nm. Fig. 4h)-i) shows color coded shape descriptor maps (derived using BioVoxxel Toolbox (Brocher, 2022, 2023) for ImageJ/FIJI (Schindelin et al., 2012)) of feret diameter (in microns) and circularity (4 *π * Area/(perimeter*perimeter)) of the "red areas" respectively. It is seen that the shape (circularity) depends of the "red areas" size (feret diameter). This is clearly visible on the plot Fig. 4j), showing derived circularity as a function of feret diameter of the "red areas". As the size of the "red areas" increases the circularity decreases, the shape changes from round to non regular and more elongated one. Additionally, two regions on the plot are clearly visible: (1) the region of "red grooves" with the mean feret diameter of 22.97 μm and mean circularity of 0.191 (non regular and more elongated structures); (2) the region of "red spots" with the mean feret diameter of 3.23 μm and mean circularity of 0.700 (more round structures).

To fully characterize the optical properties of the gold surface, the spectral reflectance (reflectance as a function of wavelength) was colorimetrically recovered for each point on the microscopic image, see Fig. 5a). This forms a hyperspectral reflectance data cube, where the reflectance R in measured for each x, y pixel position on the image, see Fig. 5d). For the reflectance recovery from measured color calibrated (D65 illuminant) image, method developed by Otsu et al. (2018). was used. The method uses data driven approach for the spectra recovery. Database of measured spectra is first separated into clusters for which a set of basis functions is precomputed by applying Principal Component Analysis (PCA). During the spectra reconstruction, the input color is first used to match to the precomputed cluster. This solves the metamerism problem, that different spectra can correspond to the same color (Otsu et al., 2018). The reflectance spectrum is reconstructed using the basis function set corresponding to the cluster. The reflectance spectrum is reconstructed in the range of 380–730 nm. The method outperforms other reconstruction methods (Smits, 1999; Meng et al., 2015) significantly by giving the lowest spectra reconstruction error (Otsu et al., 2018). The method is used here as implemented in Color Science library (Mansencal et al., 2022). It is worth to notice that since the colorimetric measurements are performed at micro meter scale, the effects of the surface roughness are visible as local changes. The amount of data and the data coverage (x,y chromaticity diagram coverage) used to derive the reconstruction is a limiting factor and main source of errors of the final reflectance reconstruction (Otsu et al., 2018). Fig. 5b) shows mean reflectance map resulted from colorimetric recovery. Regions of different reflectance ranging from 0.2 to 0.8 are visible on the surface. Fig. 5c) shows colorimetrically recovered mean gold reflectance as a function of the wavelength. The reflectance is a global average reflectance over the reflectances over the whole image region. This is compared with a measured gold surface reflectance (Bass, 1995). The reflectances were normalized to maximum value of 1 for the comparison. It is seen that the colorimetrically recovered reflectance spectrum agrees very well with the measured gold surface reflectance. To quantify the agreement between the spectra, correlation coefficient was calculated and is equal to r = 0.99182, this shows a very high similarity, which validates the reflectance recovery. This shows also the level of precision and correctness of colorimetric recovered reflectance. The visualization of the recovered hyperspectral reflectance is presented as a movie in Supporting Information.

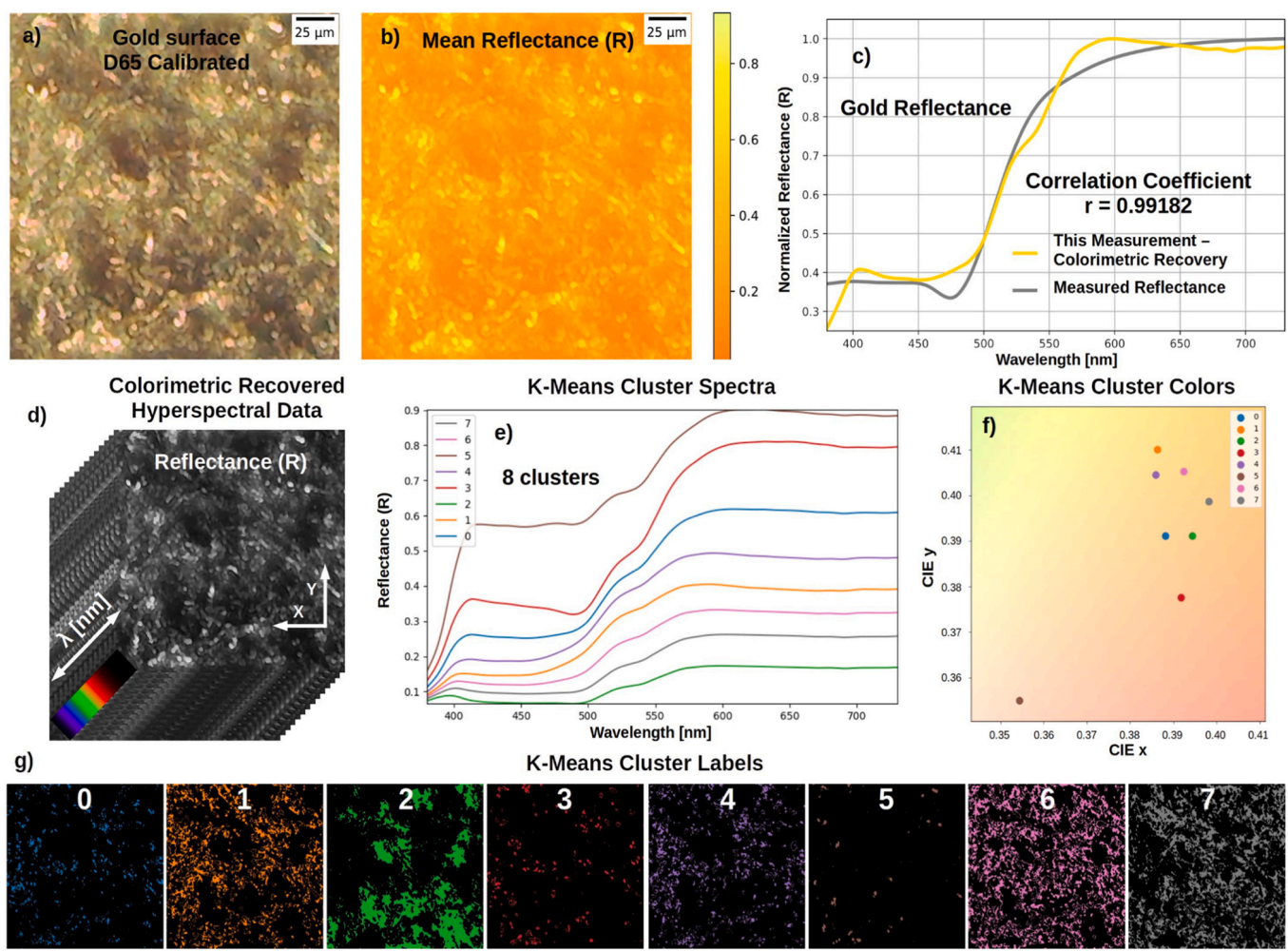

**Fig. 5.** Hyperspectral data recovery for the colorimetric microscopy (C-Microscopy) imaging of the surface of Gold 99.999% pellet at microscale. Color calibrated (D65 illuminant) optical microscopy image of Gold surface a). Colorimetrically recovered mean reflectance map b). Colorimetrically recovered normalized gold reflectance as a function of the wavelength together with measured gold reflectance spectra c). The reflectance is a global average reflectance over the whole image region. The colorimetrically recovered reflectance spectra agrees with the measured one, correlation coefficient r = 0.99182. Graphical representation of colorimetrically recovered hyperspectra reflectance data d). Results of k-means clustering into eight clusters of reflectance hyperspectral data: cluster spectra e) and corresponding cluster labels in the form of the maps. Colors of the corresponding cluster reflectance in the CIE 1931 coordinates f). It is seen that different cluster regions in microscale exhibits different type of the reflectance spectra and in consequence color. The global gold reflectance spectra is the average of the local reflectances in the microregions. The local optical properties of the surface of gold were characterized at microscale.

Supplementary material related to this article can be found online at doi:10.1016/j.micron.2023.103557.

To characterize the local changes of the reflectance spectra the hyperspectral reflectance data were clustered, grouped together into regions with similar properties by K-Means clustering using mini batch k-means method as implemented in Scikit-Learn (Pedregosa et al., 2011). The optimal number of clusters (groups) is determined using elbow method from the plot of inertia versus number of clusters, see Fig. S4 in Supporting Information. In this case the data were clustered into eight clusters. The results are presented in Fig. 5e),g), which shows cluster labels and corresponding cluster spectra. The cluster labels Fig. 5g) show eight found regions on the gold surface exhibiting different local reflectance properties. It is seen that the regions correspond also to the different local surface morphologies i.e. grooves, hills etc., when compared with Fig. 4b). Detailed morphological analysis of cluster labels by employing Minkowski Functionals analysis (Hilou et al., 2020; Armstrong et al., 2019) shows directly that different cluster exhibits different surface morphology, see Supporting Information Fig. S6. Different optical properties are directly linked with local morphology. The cluster spectra Fig. 5g) quantify the regions by showing mean reflectance spectra corresponding to each region. Looking closely into reflectances Fig. 5g) one can notice that they exhibit different dependence on the wavelength, related to different optical absorption changes. The differences between the spectra are better seen in the first derivative of the reflectance, see Supporting Information Fig. S5. These differences in consequence are manifested as different color, see Fig. 5f). The Fig. 5f) shows colors of the corresponding cluster reflectance in the CIE 1931 coordinates. It is seen that the color of the clusters changes from yellow to light red.

The studied gold surface was decomposed into microregions with different local optical properties by hyperspectral reflectance spectra clustering analysis.

This shows that at the microlevel the ordinary gold surface is not homogeneous in terms of optical response due to the presence of different surface features i.e. grooves, hills etc. This gives the possibility of tuning optical properties of gold and other different metals by intensional surface micro modification. This kind of similar structural color effects have been observed for the copper surface after modification by picosecond laser in terms of sequential color change (Peixun et al., 2013). The optical effects were only globally characterized by global reflectance at macro scale (Peixun et al., 2013). Here, the observed optical effects on gold surface are characterized at its origin at the micro level scale using C-Microscopy approach. This gives a full understanding and quantification of the local optical properties which here are the consequence of the particular surface features.

The python jupyter notebook to colorimetrically recover hyperspectral reflectance data from color calibrated (D65 illuminant) image is freely available from Zenodo (Benedykt R. Jany, 2023) (https://doi.org/10.5281/zenodo.7789585).

## 4. Conclusions

The Colorimetric Microscopy (C-Microscopy) approach using digital optical microscopy and a free software was introduced. The approach uses color calibrated (D65 illuminant) image and colorimetric calculations to obtain physically meaningful quantities i.e. dominant wavelength and excitation purity maps of the specimens at micro level scale. The systematic uncertainties related to color calibration procedure are propagated to measured quantities using Monte-Carlo calculations to gain full control of the systematic effects and to confirm the reliability of the obtain measurements. This resulted in the successful colorimetric characterization and interpretation in a physical meaningful quantitative way the collected microscopic images of the studied samples i.e. two metamorphic rocks lapis lazuli and unikite and well as the surface of pure 99.999 % gold pellet. The areas of different minerals were clearly visible for the metaphoric rocks. In the case of gold surface the effect of the red shift in the surface grooves was identified, which was not directly seen on the raw microscopic image. Later, in the C-Microscopy approach, the hyperspectral reflectance data were colorimetrically recovered. This allowed for the gold surface decomposition into microregions with different local optical properties. It was found that at the microlevel the ordinary gold surface is not homogeneous in terms of the optical response due to the presence of different surface features. This gives the possibility of tuning optical properties of metals by intentional surface micro modification. The introduced C-Microscopy approach could be used to quantify the local optical properties changes of various materials at microscale in an accessible way in contrast to the not so common microscope spectrophotometers or specialized microreflectance setups. The approach gives also the possibility to incorporate and to use the colorimetric information for the broad range of microscopic measurements. It is also worth to notice that the presented approach is not limited to microscopic images only, it could be used to analyze any type of color calibrated image. The presented Colorimetric Microscopy (C-Microscopy) approach is freely available as a set of python jupyter notebooks.

## Declaration of Competing Interest

The authors declare that they have no known competing financial interests or personal relationships that could have appeared to influence the work reported in this paper.

## Data availability

The data and the programs are available from Zenodo https://doi.org/10.5281/zenodo.7789584.

## Acknowledgments

This research was supported in part by the Excellence Initiative - Research University Program at the Jagiellonian University in Krakow.

## Appendix A. Supporting information

Supplementary data associated with this article can be found in the online version at doi:10.1016/j.micron.2023.103557.

# Supporting Information
## Quantifying Colors at Micrometer Scale
## by Colorimetric Microscopy (C-Microscopy) Approach

Benedykt R. Jany

Marian Smoluchowski Institute of Physics, Faculty of Physics, Astronomy and Applied Computer Science, Jagiellonian University, Lojasiewicza 11, 30348 Krakow, Poland

corresponding author e-mail: benedykt.jany@uj.edu.pl

| Label | Red | Green | Blue |
|---|---|---|---|
| R Squared | *0.994* | *0.997* | *0.993* |
| constant | -4.31616900 | 18.46419552 | 23.90842999 |
| r | 3.38063167 | -0.27073860 | 0.87636352 |
| g | -0.31691123 | 1.40213154 | -0.42240564 |
| b | 0.18173961 | 0.18332645 | 1.02820518 |
| r*r | -0.02464148 | 0.00229821 | -0.00724081 |
| r*g | -0.00026517 | 0.00114570 | -0.00232056 |
| r*b | 0.00007509 | 0.00018493 | -0.00071537 |
| g*g | 0.00483070 | -0.00731942 | 0.00565634 |
| g*b | -0.00456137 | 0.00097616 | -0.00006415 |
| b*b | 0.00026309 | -0.00337350 | -0.00421741 |
| r*r*r | 0.00006979 | -0.00000871 | 0.00002500 |
| r*g*b | 0.00001518 | -0.00000450 | 0.00001596 |
| g*g*g | -0.00001291 | 0.00002028 | -0.00002086 |
| b*b*b | 0.00000289 | 0.00001086 | 0.00001366 |

*Table S1: Resulting calibration coefficients for color calibration (D65 illuminant) of microscope.*

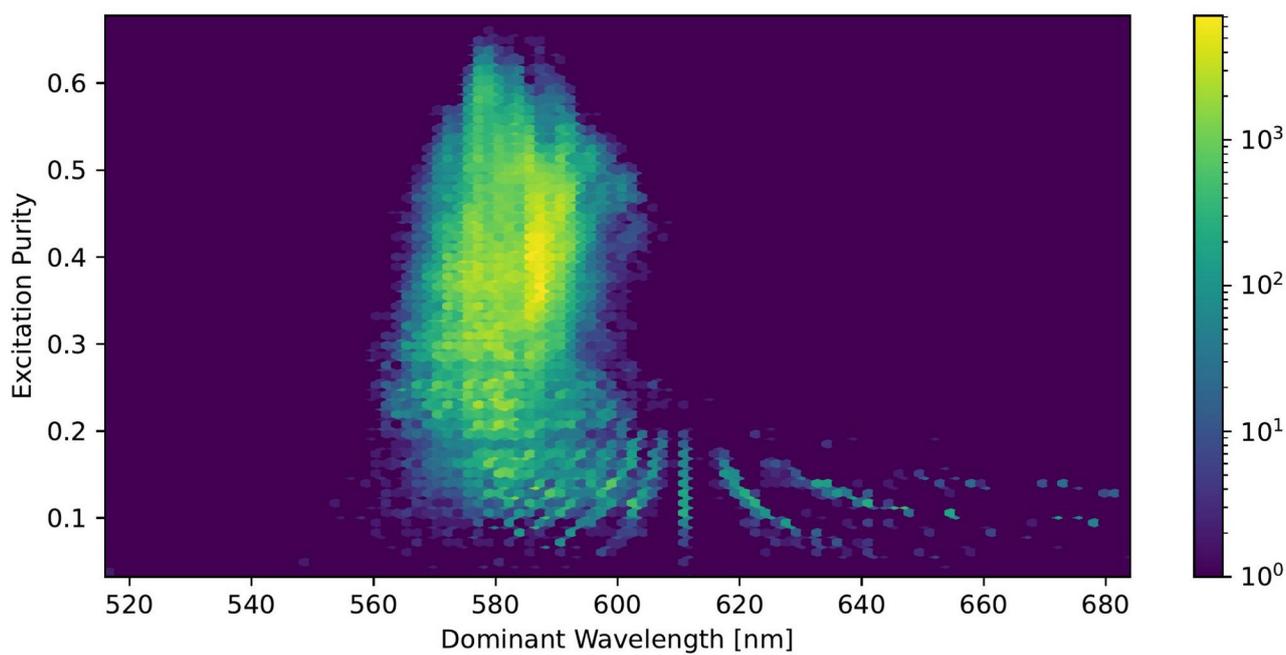

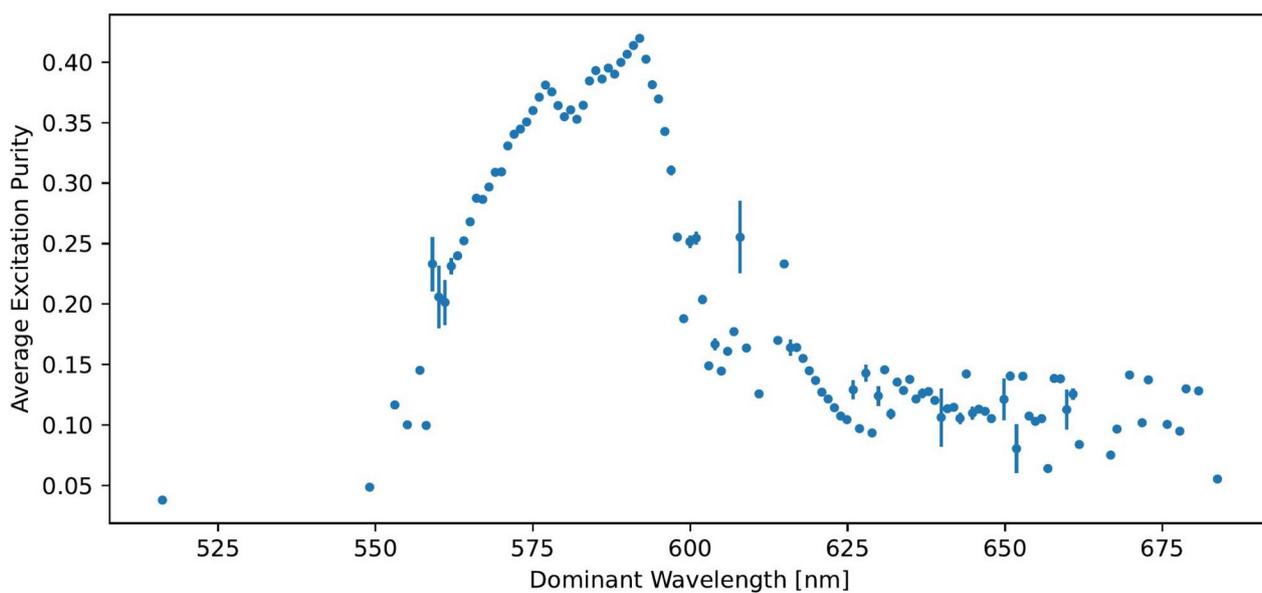

*Figure S1: Results of colorimetric microscopy (C-Microscopy) imaging of Unakite polished plate. Correlation between excitation purity and dominant wavelength (upper image). Average excitation purity as a function of dominant wavelength (lower image).*

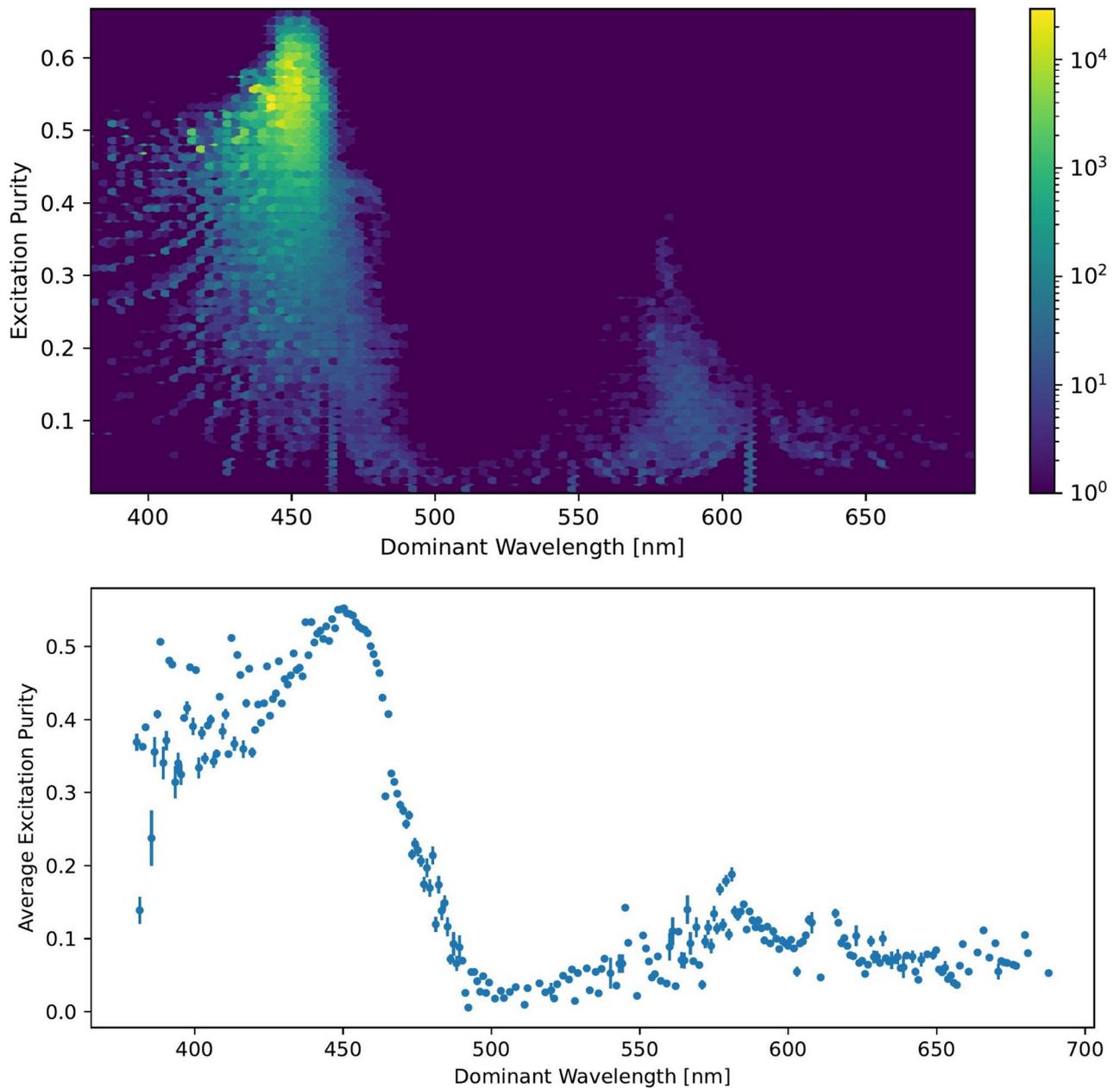

*Figure S2: Results of colorimetric microscopy (C-Microscopy) imaging of Lapis Lazuli polished plate. Correlation between excitation purity and dominant wavelength (upper image). Average excitation purity as a function of dominant wavelength (lower image).*

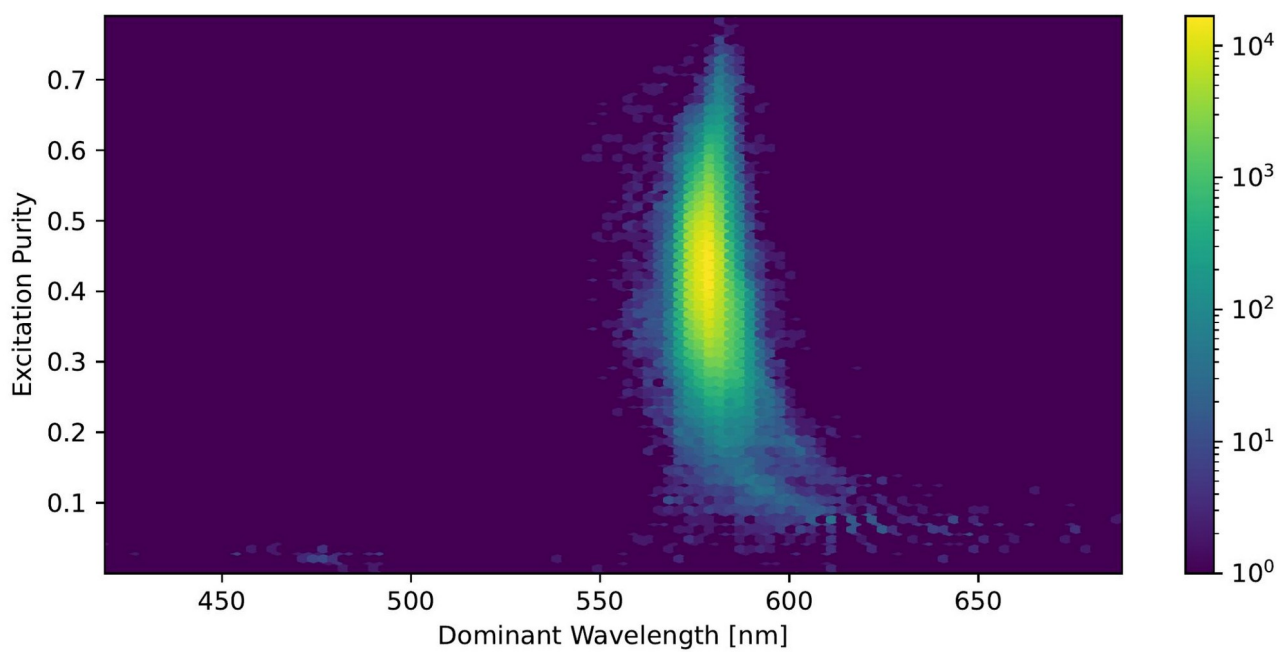
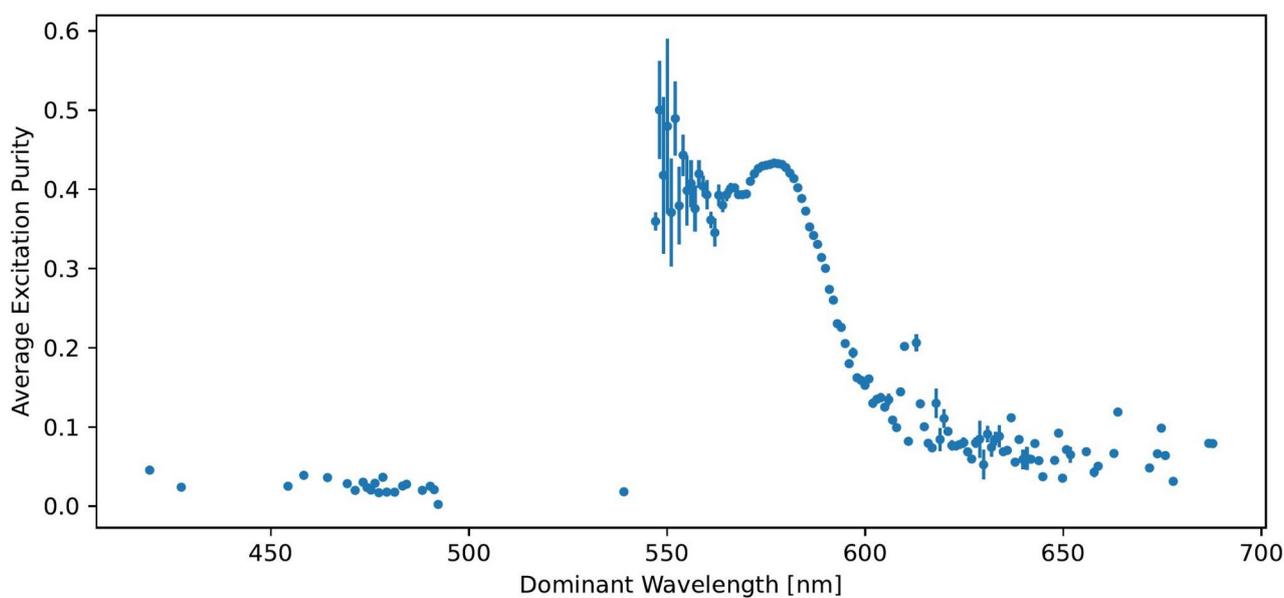

*Figure S3: Results of colorimetric microscopy (C-Microscopy) imaging of surface of Au 99.999% pellet. Correlation between excitation purity and dominant wavelength (upper image). Average excitation purity as a function of dominant wavelength (lower image).*

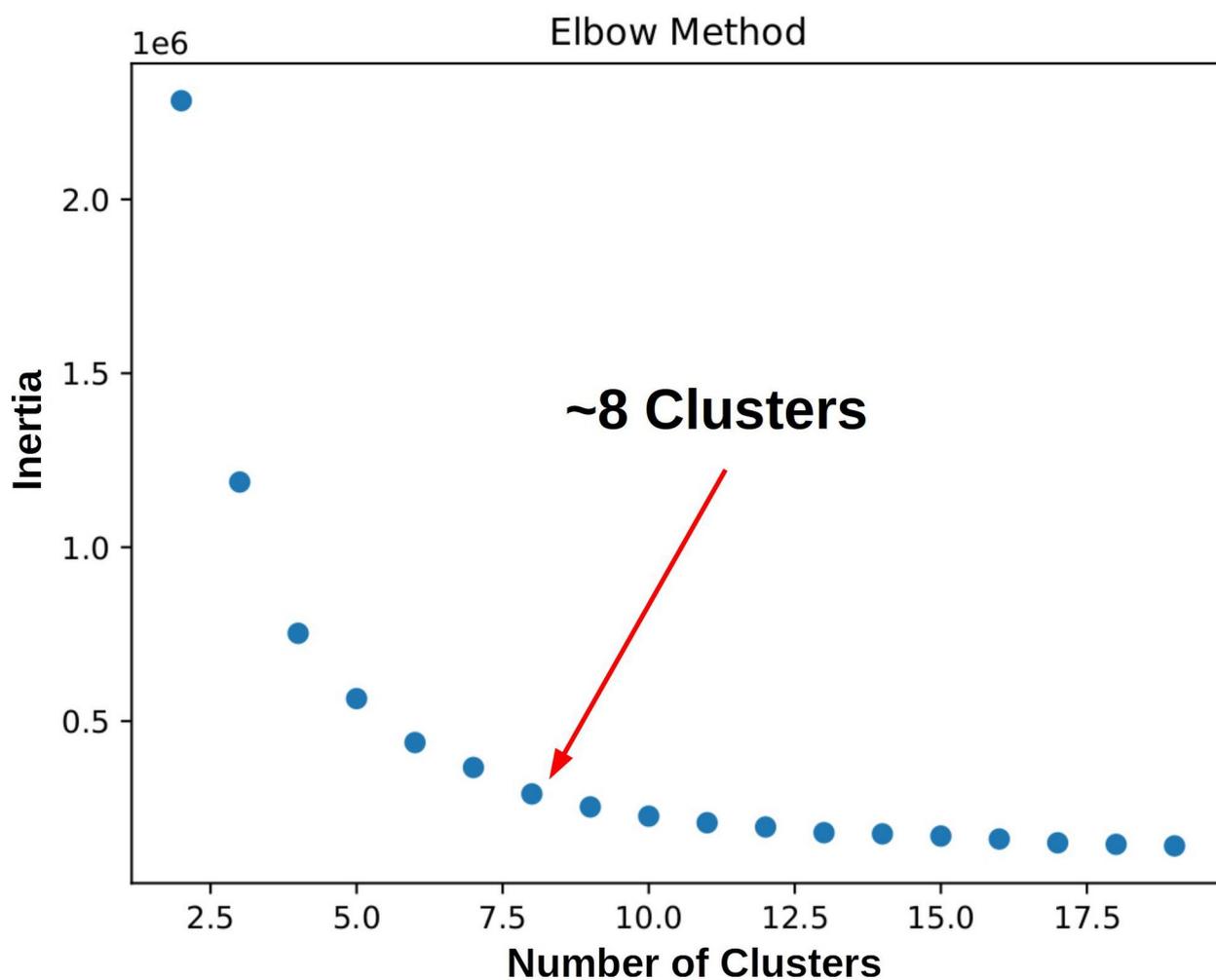

*Figure S4: Inertia versus number of clusters for mini batch k-means clustering of colorimetrically recovered hyperspectral reflectance data of gold 99.999% pellet surface at microscale. The estimated number of clusters corresponds to the 8 clusters.*

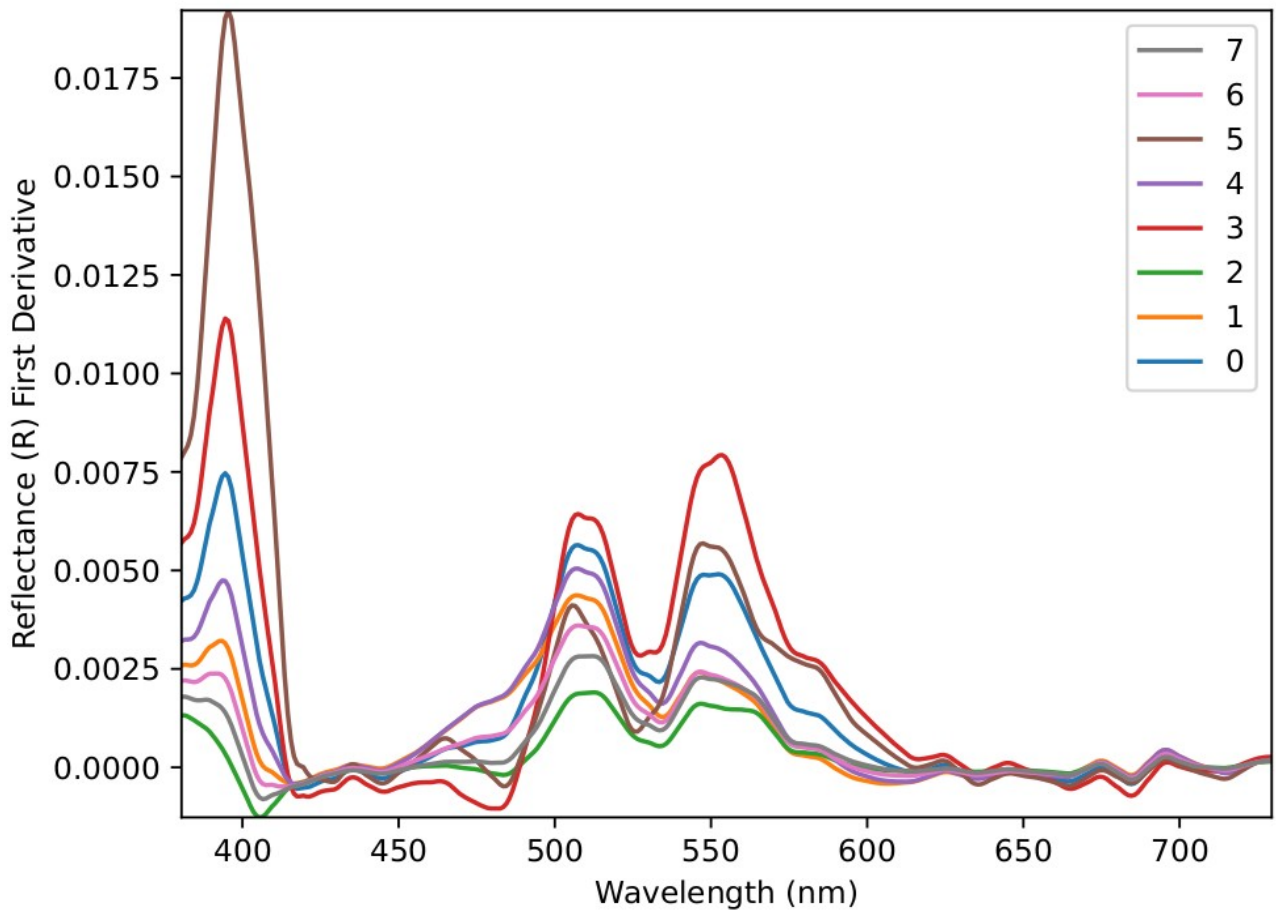

*Figure S5: Results of k-means clustering into eight clusters of reflectance hyperspectral data of the surface of Gold 99.999% pellet at microscale. First derivative of the cluster spectra i.e. reflectance spectra.*

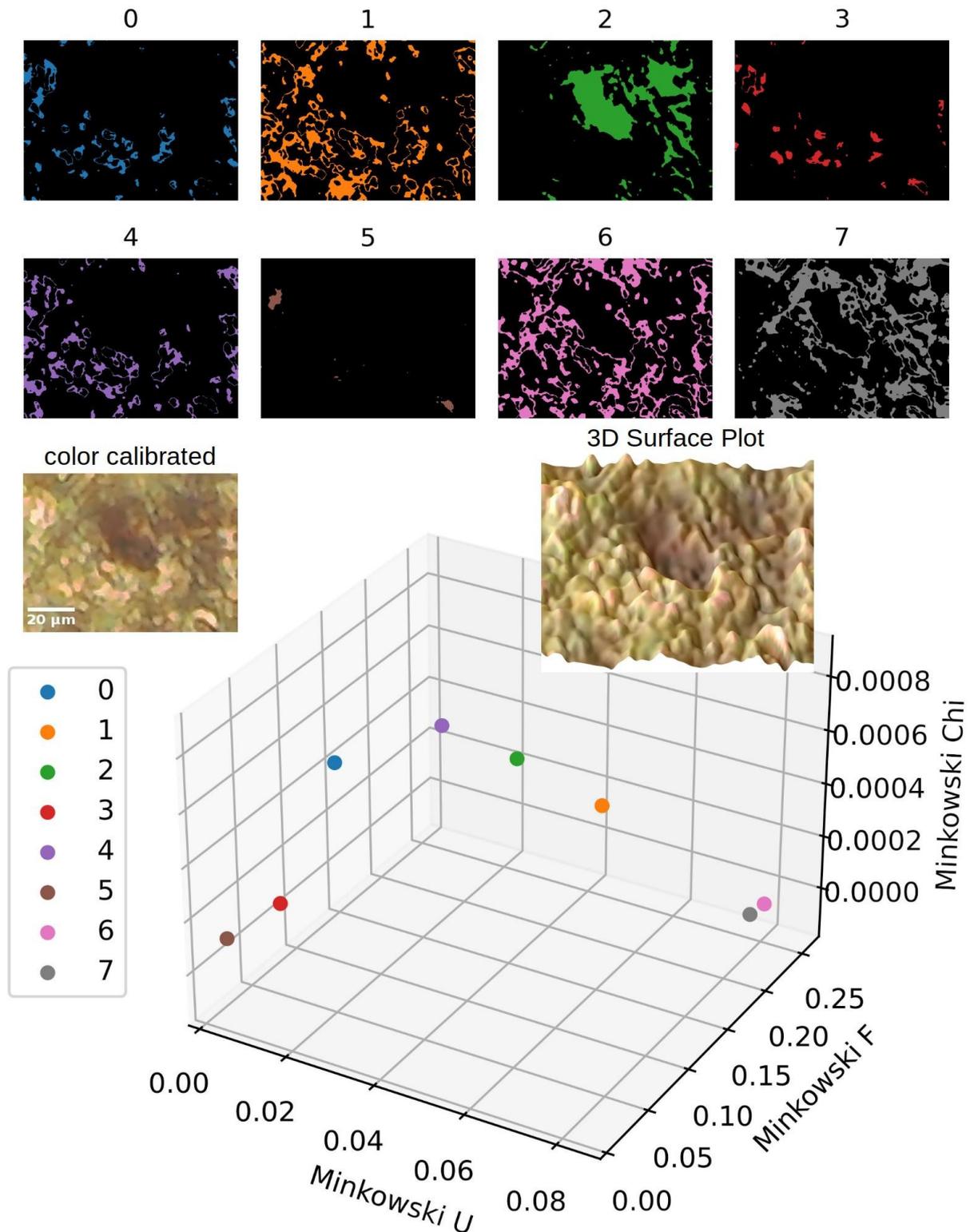

*Figure S6: Morphological analysis by Minkowski Functionals of cluster labels resulted from k-means clustering of hyperspectral reflectance of Gold 99.999% pellet surface at microscale. A close-up view (to see the details) into cluster labels together with color calibrated image of the surface and 3D surface plot (upper part). Result of Minkowski Functional morphological analysis, 3D scatter plot of Minkowski U (perimeter) vs Minkowski F (area) vs Minkowski Chi (Euler characteristic) (lower part). It is seen that different clusters (corresponding to different optical properties i.e. reflectance spectrum) exhibit different morphology in terms of Minkowski Functionals.*

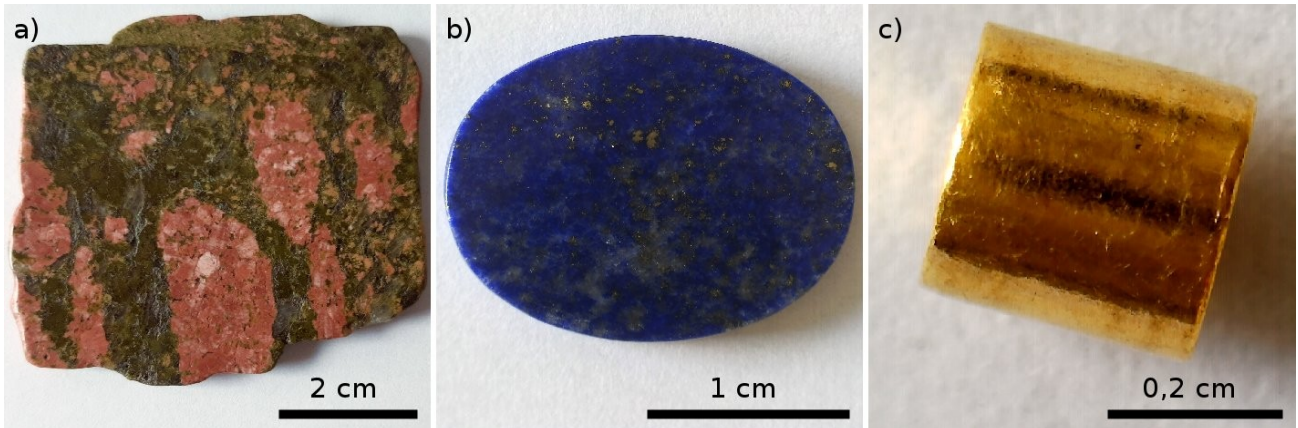

*Figure S7: Macro scale view of the measured by C-Microscopy samples: a) Unakite polished plate, b) Lapis Lazuli polished plate, c) Gold 99.999% pellet.*

| Sample Description | CIE 1931 xy coordinates D65 illuminant |
|---|---|
| Lapis Lazuli Polished Plate (This Work) average over microscopic image area | x:0.232±0.015 y:0.166±0.021 (95% CL) |
| Mineral Lazurite Powder (Lapis Lazuli) USGS Spectral Library Version 7 [1] | x:0.248 y:0.257 |
| Ultramarine (Paint Pigment) USGS Spectral Library Version 7 [1] | x:0.167 y:0.096 |
| Prussian Blue (Paint Pigment) USGS Spectral Library Version 7 [1] | x:0.189 y:0.135 |
| Cobalt Blue (Paint Pigment) USGS Spectral Library Version 7 [1] | x:0.176 y:0.167 |

*Table S2: Average color of Lapis Lazuli, as measure by C-Microscopy, in CIE 1931 xy chromaticity coordinates (D65 illuminant) in comparison to the color of Mineral Lazurite Powder and artificial blue pigments (Ultramarine, Prussian Blue, Cobalt Blue). It is seen that the color of Lapis Lazuli is different than artificial blue pigments.*

The color value, expressed as xy chromaticity coordinates (D65 illuminant), of Lapis Lazuli (Polished Plate (This Work) and Mineral Lazurite Powder) is significantly different from color of paint pigments (Ultramarine, Prussian Blue, Cobalt Blue), see Table S2. This allows for the differentiation between natural Lapis Lazuli and artificial blue pigments, thus making it possible to use C-Microscopy approach in the field of heritage science. The difference between the color value of Lapis Lazuli Polished Plate and Mineral Lazurite Powder comes from the fact that in the case of Lapis Lazuli Polished Plate the color was averaged over microscopic image area (as measured by C-Microscopy) while in the case of Mineral Lazurite Powder the color value is an average over macroscopic region. Different amounts of Lapis Lazuli constituents (blue lazurite, yellow pyrite and white calcite) in different proportions are taken for the measurements, in one case (C-Microscopy) it is a very local average in micro region in the second case there is a average over "homogeneous powder" area.